\documentclass[twocolumn,showpacs,preprintnumbers,amsmath,amssymb,10pt,a4paper]{revtex4}

\usepackage{geometry}
\usepackage{graphicx}
\usepackage{dcolumn}
\usepackage{bm}
\newcommand{\be}{\begin{equation}}
\newcommand{\ee}{\end{equation}}
\newcommand{\bd}{\begin{displaymath}}
\newcommand{\ed}{\end{displaymath}}
\newcommand{\BE}{\begin{eqnarray}}
\newcommand{\EE}{\end{eqnarray}}

\newcommand{\bx}{\ensuremath{\mathbf{x}}}
\newcommand{\by}{\ensuremath{\mathbf{y}}}
\newcommand{\bz}{\ensuremath{\mathbf{z}}}

\newcommand{\bzeta}{{\mbox{\boldmath $\zeta$}}}
\newcommand{\bDelta}{{\mbox{\boldmath $\Delta$}}}

\newcommand{\avg}[1]{\left\langle{#1}\right\rangle}

\begin{document}

\preprint{}
\title{Intrinsic noise in game dynamical learning}

\author{Tobias Galla}
\email{tobias.galla@manchester.ac.uk}

\affiliation{Theoretical Physics, School of Physics and Astronomy, The University of Manchester, Manchester M13 9PL, United Kingdom}

\date{\today}

\begin{abstract}
Demographic noise has profound effects on evolutionary and population dynamics, as well as on chemical reaction systems and models of epidemiology. Such noise is intrinsic and due to the discreteness of the dynamics in finite populations. We here show that similar noise-sustained trajectories arise in game dynamical learning, where the stochasticity has a different origin: agents sample a finite number of moves of their opponents inbetween adaptation events. The limit of infinite batches results in deterministic modified replicator equations, whereas finite sampling
leads to a stochastic dynamics. The characteristics of these fluctuations can be computed analytically using methods from statistical physics, and such noise can affect the attractors significantly, leading to noise-sustained cycling or removing periodic orbits of the standard replicator dynamics.
\end{abstract}

\pacs{02.50.Le, 87.23.Kg, 02.50.Ey, 05.40.-a}
\maketitle

Intrinsic noise has been seen to have significant effects on dynamical systems, and may alter their attractors substantially. Noise-sustained oscillations, generated via an amplification mechanism, are for example present in models of population dynamics \cite{alan}, epidemiology \cite{epidemics} or biochemical reaction systems \cite{biology}. The origin of these fluctuations is the discreteness of the dynamics in {\em finite} systems, deterministic descriptions are then no longer appropriate. The class of systems in which intrinsic noise cannot be neglected includes models of evolutionary dynamics and game theory, and much current research aims at understanding the effects of this demographic stochasticity using methods from nonequilibrium statistical mechanics and the theory of stochastic processes \cite{traulsen}.

Here, we will focus on intrinsic noise resulting from a different origin, and will consider the learning dynamics of agents in a game theoretic setting \cite{fudenberg}. This is complementary to more conventional approaches to game theory concentrating on the characterisation of equilibrium points \cite{morgenstern}, or on evolutionary processes \cite{maynard}. In the learning scenario one considers a
small number of agents who interact repeatedly in a given game, and
who observe their opponents' actions and aim to react by
adapting their own strategy profile. Such dynamical models are of
particular importance for the understanding of experiments in game
theory and behavioral economics, in which human subjects play a given
game repeatedly under controlled conditions \cite{fehr,camerer}. As a
key result we show that stochasticity, induced by imperfect sampling
of the opponents' strategy profiles, can result in trajectories quite different from those of deterministic learning,
very much akin to the mechanism by which intrinsic noise in finite
populations affects the trajectories of evolutionary systems. While the amount of intrinsic noise in evolutionary dynamics is determined by the number of individuals in the population, our objective here is to characterise the fluctuations in the learning dynamics of two fixed agents. The quantity controlling the noise strength is the number of observations made by the agents inbetween adaptation events.  Furthermore, in a deterministic setting and depending on the game, we demonstrate that memory loss can promote or impede convergence to a Nash equilibrium.

Consider a general symmetric two-player game, played
repeatedly by players $X$ and $Y$, and assume there are $p$ pure
strategies in this game. The payoff matrix is given by
$a_{ij}$ where $i,j\in\{1,\dots,p\}$. The rounds of the repeated
interaction will be labeled by $t=1,2,...$ in the following. In each
round player $X$ plays one pure strategy $i(t)\in\{1,\dots,p\}$, and
player $Y$ plays $j(t)\in\{1,\dots,p\}$. The payoff for $X$ is then
$a_{i(t)j(t)}$ and that for $Y$ is $a_{j(t)i(t)}$. If the players play
stochastically, i.e. if they resort to mixed strategies, $i(t)$ and $j(t)$ will be random variables.
Assuming that player $X$ carries a (time-dependent) mixed strategy
profile $\bx(t)=(x_1(t),\dots,x_p(t))$ and similarly
$\by(t)=(y_1(t),\dots,y_p(t))$ for player $Y$, a learning dynamics is
then a prescription used to update these strategy profiles between
subsequent rounds of the game. $x_i(t)$ here denotes the probability
with which player $X$ plays pure strategy $i\in\{1,\dots,p\}$ in round
$t$, and similarly for $y_j(t)$. Normalization requires
$\sum_{i=1}^px_i(t)=\sum_{j=1}^p y_j(t)=1$.

In order to define a specific learning dynamics, we follow
\cite{camerer,sato} and assume that each player keeps valuations
of each pure strategy, measuring their relative performance in the
past. More precisely, in a situation without memory loss, the
valuation $q_i(t)$ player $X$ has for pure strategy $i$ is the total payoff
$X$ would have obtained, had he/she always played strategy $i$ up to time $t $, and given $Y$'s actions. The valuation $r_j(t)$ player
$Y$ has for $j$ has an analogous meaning. Following
\cite{camerer,sato} players then use a logit rule  \be
x_i(t)=\frac{e^{\Gamma q_i(t)}}{\sum_{k}e^{\Gamma q_k(t)}}, ~~
y_j(t)=\frac{e^{\Gamma r_j(t)}}{\sum_{k}e^{\Gamma
r_k(t)}}\label{eq:probx}.  \ee $\Gamma\geq0$ here sets the scale of
the score valuations, and is known as the response sensitivity
\cite{camerer}.  While $\Gamma=0$ corresponds to random response,
and $\Gamma=\infty$ to deterministic play, we will here focus on the
case in which $0<\Gamma<\infty$. It is important to distinguish
between two types of randomness in the actual play:
as prescribed by (\ref{eq:probx}), the players will generally use
mixed strategies, so that their actions can be stochastic, even at given strategy valuations. Secondly, the update of the
valuations itself will contain some stochasticity as we will
detail next. We will here assume that players update their scores only once every $N$ rounds of the game, and keep them
constant inbetween. This is known as batch learning in computer science \cite{machine}. Specifically, we will assume 
\BE\label{eq:qrupdate}
q_k(t+N)&=&(1-\lambda)
q_k(t)+\frac{1}{N}\sum_{t'=t}^{t+N-1}a_{kj(t')}\nonumber \\
r_k(t+N)&=&(1-\lambda)
r_k(t)+\frac{1}{N}\sum_{t'=t}^{t+N-1}a_{ki(t')},
\EE and $q_k(t+\tau)=q_k(t)$ for all $\tau=1,2,\dots,N-1$, and
similarly for player $Y$. On-line learning \cite{machine},
i.e. updating after each round, is recovered for $N=1$. In our model all $\{q_i,r_j\}$ are updated at each adaptation event. This corresponds to reinforcement learning in which foregone payoffs are known and reinforced, equivalent to weighted fictitious play belief learning, see Ho {\em et al.} \cite{camerer}. The
interpretation of these update rules is understood best by first
considering the case $\lambda=0$: then the increment of $q_k$
between time-steps $t$ and $t+N$ is given by
$N^{-1}\sum_{t'=t}^{t+N-1}a_ {kj(t')}$. This increment is recognized as the
average payoff $X$ would have received per round had he/she played
pure strategy $k$ in all rounds $t,t+1,\dots,t+N-1$.  A non-zero value, $\lambda\in(0,1]$, accounts for memory loss.  We here note that other approaches can be taken to describe memory-loss, for example one may introduce a pre-factor $\lambda$ in the payoff terms in Eq. (\ref{eq:qrupdate}).  In this paper we follow the setup of \cite{sato}.

 The update rules are intrinsically stochastic,  we will refer to (\ref{eq:probx},\ref{eq:qrupdate}) as discrete-time stochastic learning (DTSL). After a re-scaling of time, and for large, but finite batch size $N$ we can write
\BE\label{eq:qrupdatenoise} q_k(\ell+1)&=&(1-\lambda)
 q_k(\ell)+\sum_{j}a_{kj}y_j(\ell)+\frac{\xi_k(\ell)}{\sqrt{N}}\nonumber \\
 r_k(\ell+1)&=&(1-\lambda)
 r_k(\ell)+\sum_{i} a_{ki} x_i(\ell)+\frac{\eta_k(\ell)}{\sqrt{N}},~~
\EE
where we approximate the noise variables $\xi_k,\eta_k$ as Gaussian random variables. This amounts to an expansion in $N^{-1/2}$, and within this approximation the covariances of the $\xi_k,\eta_k$ can be obtained, as we will report elsewhere \cite{galla}. In the limit of infinite batch size, $N\to\infty$, the dynamics becomes deterministic, we will refer to this as discrete-time deterministic learning (DTDL). Assuming $\Gamma\ll 1$ a continuous-time limit \cite{sato} leads to the modified replicator equations,
\BE
\dot x_i/x_i&=&\Gamma \sum_j a_{ij}y_j-\Gamma f[\bx,\by]+\lambda\sum_kx_k\ln\frac{x_k}{x_i}\nonumber\\
\dot y_j/y_j&=&\Gamma \sum_i a_{ji}x_i-\Gamma f[\by,\bx]+\lambda\sum_ky_k\ln\frac{y_k}{y_j},\label{eq:sc}
\EE
where $f[\bx,\by]=\sum_{ij}a_{ij}x_iy_j$, as previously reported and studied in \cite{sato}, see also \cite{ahmed}. This system maintains the normalisation of probabilities, and is hence $2(p-1)$-dimensional. DTDL gives rise to a discrete version of (\ref{eq:sc}). For DTSL the map is supplemented by noise. We will denote fixed-points of the noiseless map by $\bz^*=(x_1^*,\dots,x_p^*,y_1^*,\dots,y_p^*)$, they are identical to the fixed points of (\ref{eq:sc}). We now perform an expansion about the fixed point in powers of $N^{-1/2}$, akin to the expansion first proposed in \cite{vankampen}. Writing $\bz(\ell)=\bz^*+N^{-1/2}\bDelta(\ell)$, one finds
\be\label{eq:langevin}
\bDelta(\ell+1)=\mathbb{J}\bDelta(\ell)+\bzeta(\ell),
\ee
with $\mathbb{J}$ the Jacobian at the fixed-point, and where $\bzeta(\ell)$ is Gaussian white noise, with correlations among its components, which can be worked out analytically \cite{galla}. Eq. (\ref{eq:langevin}) is the discrete-time analogue of a linear Langevin equation, and the starting point for the analysis of fluctuations about the deterministic limit. In particular Eq. (\ref{eq:langevin}) allows one to compute the stationary distributions of the components of $\bDelta$, as well as their temporal correlations and power spectra $P_i(\omega)=\avg{|\widetilde\Delta_i(\omega)|^2}$, with $\widetilde\Delta_i(\omega)$ the Fourier transform of $\Delta_i(\ell)$ \cite{galla}. This follows the lines of \cite{alan}. Here we will illustrate the effects noise has on the learning dynamics using the two examples of the prisoners' dilemma, and that of the rock-papers-scissors game.

\begin{figure}[t]
\centerline{\includegraphics[width=0.48\textwidth]{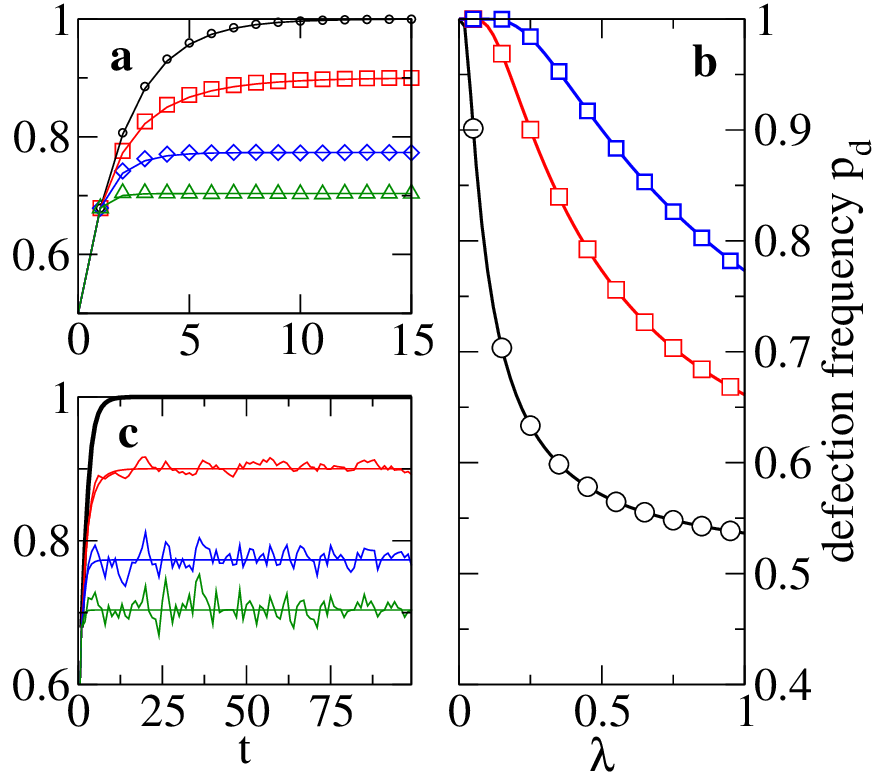}}
\caption{(Color on-line). Defection rate in the prisoners' dilemma. (a) Dynamics at $\Gamma=0.5$, $\lambda=0,0.25,0.5,0.75$ (top to bottom). Markers are from simulations of DTSL ($N=10$, averaged over $1000$ runs, defection rate shown for one fixed player), lines from DTDL; (b) Defection rate as a function of the memory-loss rate $\lambda$ for $\Gamma=1,0.5,0.1$ (top to bottom); (c) Single runs of the DTSL dynamics at $N=10$, parameters as in (a). }
\label{fig:fig1}
\end{figure}

\begin{figure}[h!!]
\vspace{2em}
\centerline{\includegraphics[width=0.48\textwidth]{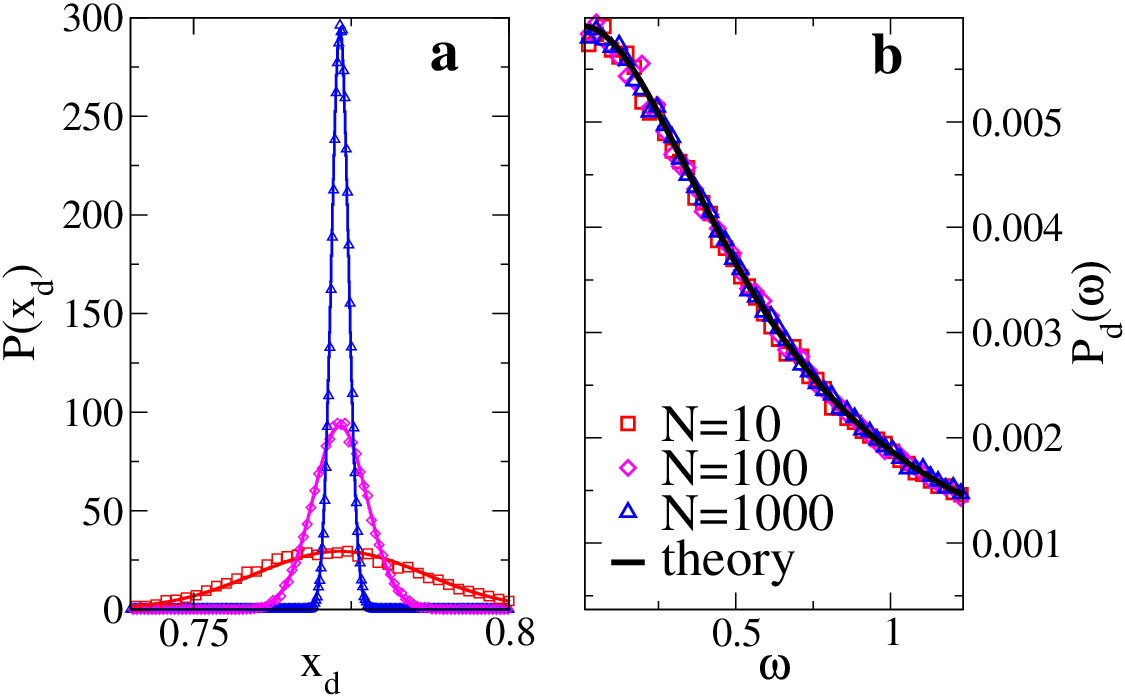}}
\vspace{1em}
\caption{(Color on-line). Defectors in the prisoners' dilemma. (a) Distribution of defection rates at $\Gamma=\lambda=0.5$, $N=1000,100,10$ from top to bottom at the peak, (b) Spectrum of fluctuations of defection rate. Symbols from simulations in both panels, solid lines from theory.}
\label{fig:fig2}
\end{figure}

The prisoner's dilemma describes a problem of mutual cooperation, where two players each face the choice whether to co-operate (C) or to defect (D). We will here choose the payoff matrix $a_{CC}=3, a_{CD}=0, a_{DC}=5, a_{DD}=1$.
The Nash equilibrium, and fixed-point of the standard replicator dynamics ($\lambda=0$) is defection, and we will in the following discuss the outcome of the batch and on-line learning dynamics with and without memory loss. As seen in Fig. \ref{fig:fig1}a, the deterministic learning dynamics converges to a fixed-point, a numerical analysis shows that this fixed-point is symmetric with respect to the exchange of players ($\bx^*=\by^*$). The defection rate of either player decreases with increasing memory loss (Fig. \ref{fig:fig1}b). The fixed-point of (\ref{eq:sc}) depends only on the ratio $\lambda/\Gamma$, and the different curves in Fig. \ref{fig:fig1}b can be collapsed. The learning dynamics at finite batch size and $\lambda>0$ yields noisy trajectories fluctuating about the deterministic mean (Fig. \ref{fig:fig1}c), averaging the noisy dynamics over independent runs reproduces the deterministic trajectory (Fig. \ref{fig:fig1}a). In Fig. \ref{fig:fig2} we address the nature of stochastic fluctuations in more detail. While deterministic learning converges towards a mixed strategy fixed point, learning at finite batch sizes leads to a distribution of mixed strategy vectors as indicated in Fig. \ref{fig:fig2}a. The width of these distributions scales as $N^{-1/2}$, and can be obtained from the theory to great accuracy. Panel \ref{fig:fig2}b demonstrates that our analytical approach captures spectral properties of the fluctuations as well, and again near perfect agreement between theory and simulations is found. These results show that the expansion in the inverse batch size is a viable analytical tool for the characterization of stochastic effects in game dynamical learning, and we will proceed to apply it to a second matrix game in the following.

\begin{figure}[t!!]\vspace{1em}
\centerline{\includegraphics[width=0.42\textwidth]{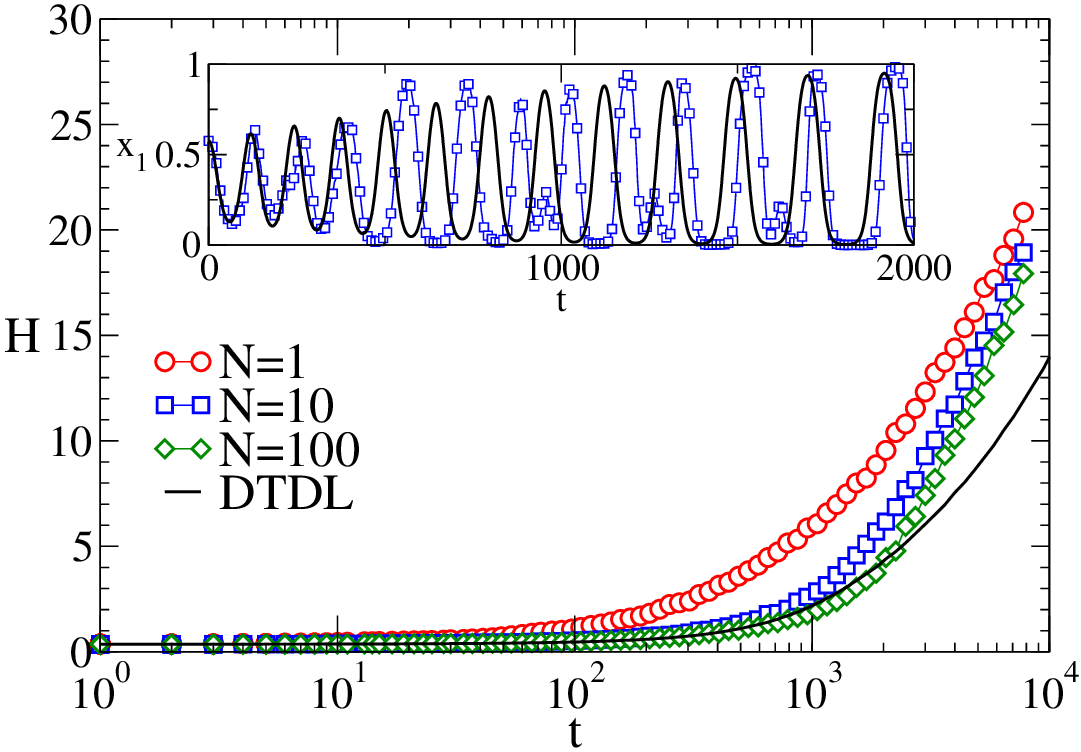}}
\vspace{0.5em}
\caption{(Color on-line). Rock-papers-scissors without memory loss ($\lambda=0, \Gamma=0.1$). Main panel shows the distance $H$ from the center of the simplex versus time. Solid line is the DTDL dynamics, markers from DTSL at finite batch size (averages over $1000$ runs). The inset shows the frequency of one of the pure strategies versus time for DTDL and for one run of DTSL, and illustrates the drift towards the edges of the strategy simplex.}
\label{fig:fig3}
\end{figure}

Rock-papers-scissors (RPS) is a game with $p=3$ strategies and cyclic dominance, as indicated by the payoff matrix $a_{RS}=a_{SP}=a_{PR}=1$, $a_{SR}=a_{PS}=a_{RP}=-1$ and $a_{RR}=a_{PP}=a_{SS}=0$. If the system is started from symmetric initial conditions, $(x_R,x_P,x_S)=(y_R,y_P,y_S)$, the continuous-time replicator dynamics, Eqs. (\ref{eq:sc}) at $\lambda=0$ reduces to a one-population dynamics, and these have one neutrally stable fixed-point at $x_R^*=x_P^*=x_S^*=1/3$, and with closed periodic orbits surrounding it \cite{gintis}. The quantity $H=-\ln(x_Rx_Px_S)-3\ln 3$ is a constant of motion \cite{gintis}, which vanishes at the neutrally stable fixed point, and indicates a measure of distance from this fixed-point. The symmetry between the two players can be broken as discussed in \cite{sato}, giving rise to the possibility of limit cycles and chaotic motion, which we do not discuss here. We first investigate the case without memory loss in Fig. \ref{fig:fig3}. The discrete-time learning dynamics at infinite and at finite batch sizes does not proceed along the cycles of the continuous-time replicator dynamics, but instead it drifts towards the edges of the strategy simplex. Fig. \ref{fig:fig3}a shows the distance $H$ from the center. This distance increases monotonically, so that the learning dynamics operates mostly at the borders of the strategy simplex after some transient time. In the deterministic case this effect is due to the discreteness in time of the learning process, the relevant eigenvalues of map at the central fixed point are given by $1-\lambda\pm i\Gamma/\sqrt{3}$, so that the fixed point is unstable for $\lambda<\lambda_c(\Gamma)=1-\sqrt{1-\Gamma^2/3}$, and stable for $\lambda>\lambda_c$. In the unstable regime fluctuations due to finite batch sizes enhance the outwards drift. 

\begin{figure}[t]\vspace{2em}
\centerline{\includegraphics[width=0.5\textwidth]{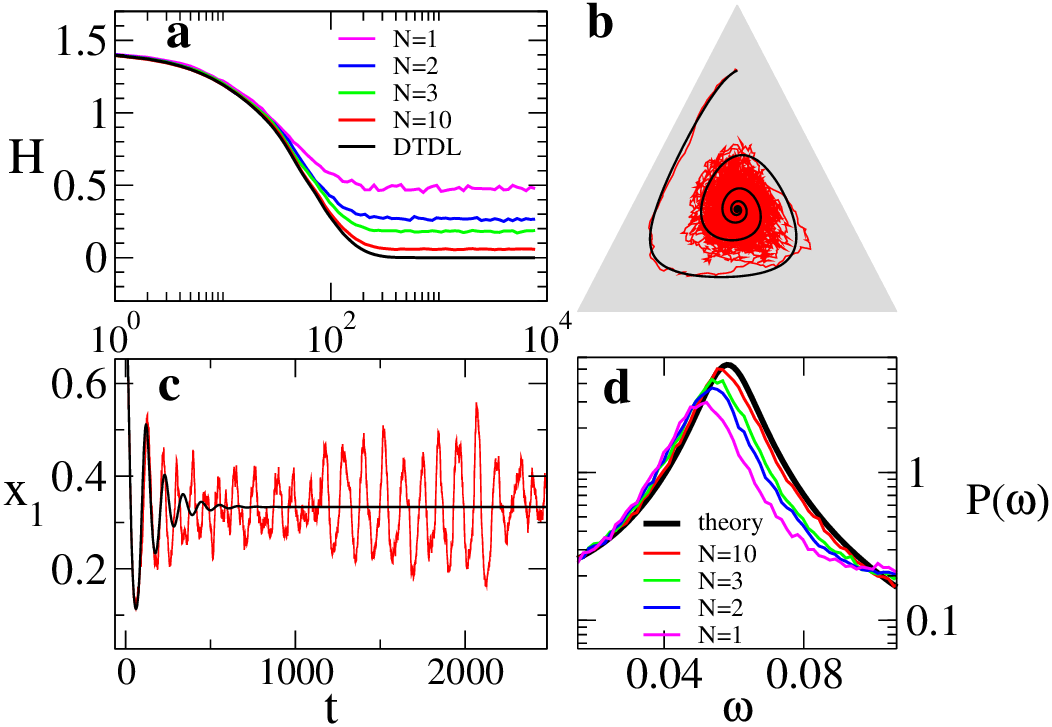}}
\caption{(Color on-line) Rock-papers-scissors at $\lambda=0.01, \Gamma=0.1$. (a) Distance $H$ versus time; (b) deterministic and stochastic trajectories ($N=10$) in the strategy simplex; (c) probability of playing rock for the same run as in (b); (d) power spectra of fluctuations for $N=1,2,3,10$ compared to theory. }
\label{fig:fig4}
\end{figure}

The differences between the noise-free learning process and on-line adaptation for the case $\lambda>\lambda_c$ is studied in Fig. \ref{fig:fig4}. Here the fixed point of the DTDL dynamics is stable. The eigenvalues of the Jacobian $\mathbb{J}$ at the fixed point are complex, and hence a resonant amplification of fluctuations is possible similar to the enhanced demographic fluctuations reported in \cite{alan}. Indeed, Fig. \ref{fig:fig4} shows that the stochastic learning dynamics at finite batch size sustains coherent stochastic oscillations about the deterministic fixed-point. Their power spectrum can be computed based on an analysis of Eq. (\ref{eq:langevin}). Results are compared with simulations in Fig. \ref{fig:fig4}d, and as seen the agreement is excellent, provided the batch size is large enough to justify the expansion in $N^{-1/2}$. Fig. \ref{fig:fig4} shows that this is the case even for small batch sizes, for other games this will most likely depend on the number of strategies available to the players. These phenomena are dynamically similar to those in evolutionary systems, where a linear scaling of extinction times in the system size have been reported for neutrally stable dynamics \cite{traulsen}. In the learning system there is no extinction, but escape times from a region around the fixed point can be measured \cite{galla}, and a similar linear scaling in the batch size is found for the neutrally stable case $\lambda=\lambda_c$. In the stable phase escape is sub-extensive, in the unstable regime escape times grow faster than linearly in $N$, very akin to what is reported in \cite{traulsen}.

Fluctuations in finite populations have profound consequences in evolutionary game theory, and we have here shown that similar stochastic effects can be seen in a learning-theoretic scenario. The source of noise is different from that in evolutionary systems, and the analogue of finite populations are finite batches of observations which players make inbetween adaptation events. Our analysis demonstrates that memory loss can lead the system away from Nash equilibria and bring about co-operation in social dilemmas. In cyclic games such as RPS convergence is only possible with sufficient memory loss, the center of the strategy simplex then becomes a stable fixed point for deterministic learning. The stochasticity and discreteness in the adaptation dynamics can affect the asymptotic attractors considerably, and noise-sustained oscillations can be observed. These oscillations are induced by an amplification mechanism similar to that observed in population dynamics \cite{alan} and in other biological systems, and may have significant amplitudes impeding the convergence to the Nash equilibrium. We expect this to be the case for a variety of different games and learning algorithms \cite{galla}, with compelling consequences for the learnability of games and their Nash equilibria. Deterministic learning of asymmetric games is known to lead to chaotic motion \cite{sato}, and we expect that a dynamics with imperfect sampling would make it even less likely that the players collectively retrieve a Nash equilibrium.

The author thanks J. D. Farmer for discussions, and Research Councils UK for financial support.

\end{document}